\documentstyle[epsf]{article}
\epsfverbosetrue

\font\tenrm=cmr10
\font\tenit=cmti10
\font\elevenbf=cmbx10 scaled\magstep 1
\font\elevenrm=cmr10 scaled\magstep 1
\font\elevenit=cmti10 scaled\magstep 1

\renewenvironment{thebibliography}[1]
 { \elevenrm
   \begin{list}{\arabic{enumi}.}
    {\usecounter{enumi} \setlength{\parsep}{0pt}
     \setlength{\itemsep}{3pt} \settowidth{\labelwidth}{#1.}
     \sloppy
    }}{\end{list}}

\def\gtap{\ \raisebox{-.4ex}{\rlap{$\sim$}} \raisebox{.4ex}{$>$}\ }
\def\beq{\begin{equation}}
\def\eeq{\end{equation}}
\def\bea{\begin{eqnarray}}
\def\eea{\end{eqnarray}}
\newcommand{\tb}{\tan \beta}
\newcommand{\wti}{\widetilde}

\newcommand{\bsgamma}{b\to s \gamma}

\newcommand{\thb}{t\to H^+b}
\newcommand{\tstopneu}{t\to {{\wti u}_1} {{\wti \chi}^0}}
\newcommand{\nn}{\nonumber}
\def\plb#1#2#3{    {\elevenit Phys. Lett. }{\elevenbf B#1} (19#2) #3}
\def\prd#1#2#3{    {\elevenit Phys. Rev. }{\elevenbf D#1} (19#2) #3}

\def\prl#1#2#3{    {\elevenit Phys. Rev. Lett. }{\elevenbf #1} (19#2) #3}

\begin{document}
\begin{center}
\vglue 0.6cm
{{\elevenbf  RARE SUPERSYMMETRIC TOP QUARK DECAYS     \\}
%
\vglue 1.0cm
{\tenrm FRANCESCA M.~BORZUMATI \\}
\baselineskip=13pt
{\tenit  II.\ Institut f\"ur Theoretische Physik
\footnote{Supported by the Bundesministerium f\"ur Forschung und
 Technologie, 05 5HH 91P(8), Bonn, FRG.   },
            Universit\"at Hamburg \\}
\baselineskip=12pt
{\tenit 22761 Hamburg, Germany \\} }
\vglue 0.8cm
{\tenrm ABSTRACT}

\end{center}

\vglue 0.3cm
{\rightskip=3pc
 \leftskip=3pc
 \tenrm\baselineskip=12pt
 \noindent
Two supersymmetric decays of the top quark, $\thb$ and $\tstopneu$,
are discussed within the framework of the Minimal Supersymmetric
Standard Model with radiatively induced breaking of
$SU(2)\times U(1)$. The present possibility of detecting these decays,
given the available bounds on supersymmetric parameters,
is compared with the situation a Next $e^+e^-$ Linear Collider would
face if supersymmetric particles were still undiscovered after
LEP~II. The indirect implications for $\thb$ and $\tstopneu$
of a possible detection of the bottom quark decay $\bsgamma$ at the
Standard Model level are taken into account.  }

\vglue 0.6cm
{\elevenbf\noindent 1. Introduction}
\vglue 0.4cm
\baselineskip=14pt
\elevenrm
Among the rare supersymmetric decays of the top quark, subject of this
talk, I have chosen to discuss only two decays which can be
far from rare~\cite{EE92}: the two decays into charged
Higgs plus bottom, $\thb$, and into the supersymmetric partner of
the top or stop, ${\wti u}_1$, plus neutralino, $\tstopneu$.
I have left out Flavour Changing Neutral Current (FCNC) decays
such as $t\to c \gamma$,~$t \to c Z$~and~$t \to c h^0$, which
negligible in the Standard Model (SM)~\cite{EE92}, cannot be brought
up to detectable levels in supersymmetry.
Non-trivial enhancements for the decay $t \to c h^0$ (and more modest
ones for $t\to c \gamma$,~$t \to c Z$~\cite{LUKESAVAGE})
can be obtained in a non-supersymmetric context,
with the SM Higgs sector extended in a
non-conventional way to contain two doublets. This topic is subject of
an independent contribution to this conference~\cite{GEORGE}.

Although much had been said about $\thb$~and~$\tstopneu$ in
the past (see references in~\cite{MEHAMBURG}), no investigation had
been performed to find out whether the
ranges of masses needed for these decays to be kinematically allowed,
are actually present in
realistic supersymmetric models. Worrisome indications about rather
heavy charged Higgs $H^\pm$ to be expected within the Minimal
Supersymmetric Standard Model (MSSM) have been recently brought
up~\cite{OLECHPOK}. Moreover,
discussions have been
triggered by refs.~\cite{HEWBAGG} about the possibility of having the
channel $\thb$ closed by the observation of the
bottom quark decay $\bsgamma$ with
a Branching Ratio compatible with the SM prediction.

\begin{figure}[t]
\epsfxsize=7.6 cm
\leavevmode
\epsfbox[60 360 523 751]{thbm150_dat1.ps}
\epsfxsize=7.6 cm
\epsfbox[60 360 523 751]{thbm150_dat2.ps}
\caption[f5]{\small{Allowed values of $m_{{H^-}}$ for the decay
 $\thb$   }}
\label{beforebsg}
\end{figure}

\looseness=-1
Some work has been done to investigate the prospects for these
two top quark decays at the moment, given the present limits on
supersymmetric parameters coming from LEP~I and the TEVATRON, and in
the situation that
supersymmetric particles are still undiscovered after
LEP~II~\cite{MEAGAIN,MEHAMBURG}.
The framework considered is the MSSM with spontaneous breaking of
the gauge group $SU(2)\times U(1)$ induced by renormalization effects
of the mass parameters appearing in the tree-level potential of the
neutral Higgs sector of this model. The embedding in a generic
grand-unified scheme is also considered.
Details regarding the procedure followed in this search
and the approximations made are given
in~\cite{MEAGAIN}, while a more thorough discussion of the results
presented here can be found in~\cite{MEHAMBURG}.

The experimental situation which the Next Linear $e^+e^-$ Collider
would have to face after LEP~I and LEP~II is
mimicked by imposing
different lower bounds to the supersymmetric particles:
gluinos $\wti g$, charginos ${\wti \chi}^-$ (the lightest is
conventionally denoted as ${{\wti \chi_2}^-}$),
neutralinos ${\wti \chi}^0$, charged and neutral
sleptons $\wti l$,~$\wti \nu$, up- and
down-squarks $\wti u$,~$\wti d$,
and neutral Higgses (with $h_2^0$ the lightest of the two
CP-even states). I shall refer to these two choices of bounds as
SET~I and SET~II. They are specifically given by:
\bea
                          &      &
m_{\wti g}            \ \     >   120\,{\rm GeV},   \quad\quad
m_{{\wti d_1},{\wti u_2}}     >   100\,{\rm GeV},   \quad\quad
m_{\wti u_1}                  >  \ 45\,{\rm GeV},   \qquad\quad  \nn \\
& {\rm SET \quad I\phantom{I}}:\quad   &
m_{\wti\chi^-_2}              >\ \ 45\,{\rm GeV},   \quad\quad
m_{\wti\nu_1}         \  \  \,>\,\ 45\,{\rm GeV},   \quad\quad
m_{\wti l_1}                \,>\   45\,{\rm GeV},   \qquad\quad  \nn \\
                          &      &
m_{h_2^0}                   \,>\ \ 30\,{\rm GeV},   \quad\quad
m_{\wti\chi^0_1}      \  \  \,>\,\ 20\,{\rm GeV},   \qquad\quad
\eea
\bea
                          &      &
m_{\wti g}            \ \     >   140\,{\rm GeV},   \quad\quad
m_{{\wti d_1},{\wti u_2}}     >   120\,{\rm GeV},   \quad\quad
m_{\wti u_1}                  >  \ 80\,{\rm GeV},   \qquad\quad  \nn \\
& {\rm SET \quad II}:\quad       &
m_{\wti\chi^-_2}              >\ \ 80\,{\rm GeV},   \quad\quad
m_{\wti\nu_1}        \  \   \,>\,\ 80\,{\rm GeV},   \quad\quad
m_{\wti l_1}                \,>\   80\,{\rm GeV},   \qquad\quad  \nn \\
                          &      &
m_{h_2^0}                   \,>\ \ 70\,{\rm GeV},   \quad\quad
m_{\wti\chi^0_1}     \  \   \,>\,\ 40\,{\rm GeV},   \qquad\quad
\eea
where the limit on $m_{{\wti \chi_1}^0}$ in SET~II is 
induced by the limit on $m_{{\wti \chi_2}^-}$. The present limits from
the TEVATRON are kept into account in SET~I (in the way considered
more appropriate for the MSSM) and a mild improvement
is assumed for SET~II; see~\cite{MEHAMBURG}.

The results shown in this talk are limited to $m_t=150\,{\rm GeV}$, but
similar features are obtained for different values of $m_t$.

\newpage
{\elevenbf\noindent 2. Decay $\thb$}
\vglue 0.4cm
A systematic study of the $(m,M)$ supersymmetric parameter space in
the range
$0\leq m \leq 500\,{\rm GeV}$,~$-250 \leq M \leq 250\,{\rm GeV}$ for
several values of $\tb$, between $3$ and $35$~\footnote{for the
rationale behind this choice of values for $\tan \beta$,
see~\protect\cite{MEAGAIN,MEHAMBURG}},
leads to the
values of masses for the charged Higgs $H^-$ shown in
fig.~\ref{beforebsg}. When SET~I of lower bounds is imposed, masses
as small as $90\,{\rm GeV}$ are obtained for the highest values of
$\tb$ considered (from $28$ to $35$). The
available phase space (the running mass $m_b(M_Z)$ is here
$\sim 3.6\,{\rm GeV}$) shrinks rapidly when $\tb $ decreases,
reaching a minimum around $\tb =9$, for which the smallest
$m_{H^-}$ obtained is $\sim 120\,{\rm GeV}$. The minimum allowed value of
$m_{H^-}$ tends to increase
again up to $m_{H^-} \sim 100\,{\rm GeV}$ when $\tb$ is reduced from
$9$ to $3$.

\begin{figure}[h]
\begin{center}
\epsfxsize=7.6 cm
\leavevmode
\epsfbox[60 360 523 751]{www.ps}
\end{center}
\caption[f5]{\small $\tb$ dependence of the ratio $R_H$}
\label{ratio}
\end{figure}
Not visible in this figure is the fact that
the low values of masses obtained for $\tb\!=\!3$ are less
``probable'' than for high $\tb$, since obtained in much
smaller regions of the same $(m,M)$ parameter space considered;%
{}~see~\cite{MEAGAIN}.

The results described here differ from the ones
presented in~\cite{OLECHPOK}. The disagreement may be due to
differences in the assumptions underlying the two 
calculations,
al\-though the precise reason is not yet clear.

A strong dependence on $\tb $ is present also in the expression for
the width $\Gamma (\thb)$. The ratio
$R_H\equiv \Gamma(\thb)/\Gamma(t \to W^+ b)$ plotted in
fig.~\ref{ratio} versus $\tb$ for three different values of $m_{H^-}$,
points to the intermediate values of $\tb$ as to the disfavored ones.

\begin{figure}[tb]
\begin{center}
\epsfxsize=7.6 cm
\leavevmode
\epsfbox[60 360 523 751]{wdhb.ps}
\end{center}
\caption[f5]{\small Ratio $R_H$ for the points 
in fig.~\ref{beforebsg},~SET~I}
\label{wdhb}
\end{figure}
The overall $(m_{H^-},\tb)$ dependence of this ratio
for the phase space of fig.~\ref{beforebsg}, SET~I,
is shown in fig.~\ref{wdhb}.

The value $\tb= 3$ yields a maximum ratio $R_H$ of about $5\%$;
intermediate values,
$3\!<\!\tb\!<\!15$, are doubly penalized
by the intrinsic drop in the rate shown in fig.~\ref{ratio} and the fact
that the masses $m_{H^-}$ obtained in these cases are heavier;
values of $\tb$ greater than $25$ can give ratios $R_H$ as big as
$20\!-\!30\%$.
The subsequent decay of $H^-$ into the pair $\tau,\nu_\tau$, with a
rate practically equal to one (for the values of $\tb$ considered
here) contributes to making the prospects for the detection of this
decay mode rather
optimistic; for a discussion on this issue see~\cite{VENTURI}.

\begin{figure}
\begin{center}
\epsfxsize=7.6 cm
\leavevmode
\epsfbox[60 360 523 751]{thbm150_smrs1.ps}
\epsfxsize=7.6 cm
\epsfbox[60 360 523 751]{thbm150_smrs2.ps}
\end{center}
\caption[f5]{\small{What remains of fig.~\ref{beforebsg} after
    imposing the constraints coming from $\bsgamma$   }}
\label{afterbsg}
\end{figure}
However, one has to be aware of the fact that an increase of the lower
bounds of supersymmetric particles up to the values of SET~II
drastically reduces the phase space available for this decay, as shown
in fig.~\ref{beforebsg}. The expected rates, which can be read
with some effort from fig.~\ref{wdhb}, are still far from small.

A last check has been performed~\cite{MEAGAIN,MEHAMBURG} to verify
whether the indirect constraints on the supersymmetric parameter space
due to the expected, imminent detection of the decay $\bsgamma$ at the
SM level can be strong enough to kinematically forbid this decay mode.
The ranges of masses $m_{H^-}$ one is left with, after imposing that
the values of $BR(\bsgamma)$ in the MSSM be limited to the interval
$2.9\!-\!4.8\times 10^{-4}$ (the range of allowed values of
$BR(\bsgamma)$ within the SM for the same value of
$m_t$~\cite{MEAGAIN}), are shown in fig.~\ref{afterbsg} for the two
sets of lower bounds, SET~I,~SET~II\footnote{The results in
fig.~\ref{afterbsg}, for SET~I, are slightly more optimistic than those
shown at the conference. The lower values of $\tb$ were then
found to be forbidden, due to
the neglect of some subleading contributions to $BR(\bsgamma)$.
However, even now, the density of points obtained in these cases is
still rather small.}.

\vglue 0.5cm
{\elevenbf \noindent 3. Decay $\tstopneu$}
\vglue 0.4cm
\begin{figure}[tb]
\begin{center}
\epsfxsize=9.68 cm
\leavevmode
\epsfbox[18 140 534 732]{stneu_new.ps}
\end{center}
\caption[f5]{\small Allowed phase space for the decay $\tstopneu$}
\label{stneu}
\end{figure}

The presence of large left-right entries in the up-squark mass matrix
is such to allow one light mass eigenstate ${\wti u}_1$,
or stop,
in principle, much lighter than
the remaining squarks. Moreover, also the mass of the lightest
neutralino, ${{\wti \chi}_1}^0$ can be rather light within the
MSSM. An investigation of the 2-dimensional parameter space $(m,M)$
for six values of $\tb$,~i.e.~$3,9,15,20,25,30$, leads to the results
shown in fig.~\ref{stneu} for the allowed phase space of the decay
$\tstopneu$, when the set of lower bounds SET~I is considered.
As in the case of $\thb$, the region of
parameter space studied is limited to the range
$0\leq m \leq 500\,{\rm GeV}$,~$-250 \leq M \leq 250\,{\rm GeV}$,
except for $\tb\!=\!30$. In this case, in fact, viable masses
$m_{{\wti u}_1}$,~$m_{{\wti \chi_1}^0}$ are obtained for $m$ and
$\vert M \vert$ as big as $700\,{\rm GeV}$ and $350\,{\rm GeV}$,
respectively.

Also for this decay,
the size of the allowed phase space
decreases for increasing values of $\tb$, reaches a minimum, and
starts increasing again after $\tb=15$.
The dependence on $\tb$ of the width $\Gamma(\tstopneu)$ is
less pronounced than for the decay $\thb$. The rates obtained for all
values of
$\tb$ considered in this case, are plotted in
fig.~\ref{gone}: they can be $\sim 10\%$ for the lowest allowed values
of $m_{{\wti u}_1}$.

The produced ${\wti u}_1$ can then decay 
as ${\wti u}_1 \to {\wti \chi}_1^+ b$,
if $m_{\wti \chi_1} < m_{{\wti u}_1}$.
This is the case for all the
points in fig.~\ref{stneu} located below the solid lines,
which practically coincide, except for $\tb=3,30$,
with the contours of the regions 
obtained.

The alternatives, when $m_{\wti \chi_1} > m_{{\wti u}_1}$,
for $\tb = 3,30$ and $25$ (in a tiny corner at the lower
values of $m_{{\wti u}_1}$), are, in principle, the three-body decays
mediated by virtual charginos,
${\wti u}_1 \to b l {\wti \nu}_1$,~$b {\wti l}_1 \nu$,
and ${\wti u}_1 \to b W^+ {\wti \chi}^0$,
mediated by a virtual chargino and/or
a down-squark.

\begin{figure}[tb]
\begin{center}
\epsfxsize=7.6 cm
\leavevmode
\epsfbox[60 360 523 751]{gstn_one.ps}
\end{center}
\caption[f5]{\small Rates for the decay $\tstopneu$}
\label{gone}
\end{figure}
However, given the values of
$m_{{\wti u}_1}$, $m_{{\wti \chi_1}^0}$ in the points above the solid
lines, it is clear that this last possibility can only occur with
off-shell $W$-bosons. Mo\-reo\-ver,
an explicit check of the values of masses which sleptons
have in the regions of phase space where
$m_{{\wti \chi_2}^-} > m_{{\wti u}_1}$, leads to the conclusion that
also ${\wti u}_1 \to b l {\wti \nu}_1$, $b {\wti l}_1 \nu$ occur
as four-body decay, with virtual sleptons producing leptons plus
neutralinos. The prospects of sizeable decay rates for stop lighter
than
charginos look, therefore, rather grim. However, the MSSM provides
another interesting possibility: the two-body decay mode
$ {\wti u}_1 \to c {\wti \chi_1}^0$, a FCNC decay due to the
coupling ${\wti \chi}^0-u_i-{\wti u_j}$ ($i\neq j$) induced through
renormalization effects by the soft supersymmetry-breaking terms.
In spite of the small coupling, this decay mode is clearly winning
when compared to the four-body decay channels, which suffer from
severe phase space suppressions~\cite{JAP}.

It is interesting to observe that the
two decays $\thb$,~$\tstopneu$, are sensitive to different regions
of the supersymmetric parameter space.
The highest rates for $\tstopneu$ can be obtained in regions
were $m_{H^-}$ is already far too big for
$\thb$ to be kinematically allowed,~see~\cite{MEHAMBURG}. Therefore, in
the case of a heavy charged Higgs, the decay $\tstopneu$ would,
in principle, play the role of the ``golden'' top-decay-candidate, if
all the other supersymmetric masses could be frozen at the values
obtained in this search. However, an increase in $m_{{H^-}}$ brings
up with itself also other masses. The reduction of the
available phase space when the lower bounds SET~II are applied, is as
severe as the one suffered by $\thb$, with only two tiny regions of
points remaining for $\tb=3$ and $30$. The ratio
$\Gamma(\tstopneu)/\Gamma(t\to W^+b)$, though, can still
reach $5-6\%$, even in points where $m_{{H^-}}>>150\,{\rm GeV}$.

Bad news for this decay mode come from the indirect constraints which may
be imposed by the detection of $\bsgamma$ at the SM level. Few points
of the phase space shown in fig.~\ref{stneu}, for SET~I, remain. They
all are at values of $m_{{\wti u}_1} \gtap 80\,{\rm GeV}$. However,
the possibility of detecting $\tstopneu$ seems to be completely closed
if the $\bsgamma$ contraints are implemented on the phase space available
when the lower limits SET~II are imposed.

\vglue 0.5cm
{\elevenbf \noindent 5. Acknowledgements \hfil}
\vglue 0.3cm
It is a pleasure to thank the organizers of this conference and the
colleagues of the Top Working Group for the Workshop on $e^+e^-$
Linear Collider, M\"unchen, Annecy, Hamburg, 1992-1993. Special thanks
go to P.~Igo-Kemenes and P.~Zerwas and gratitude to Z.~Jakubowski
for help in the use of fancy figure style files.
\vglue 0.5cm
{\elevenbf\noindent 6. References \hfil}
\vglue 0.3cm

%
\end{document}